\documentclass[preprint]{elsarticle}

\usepackage{amsmath}
\usepackage{amssymb}
\usepackage{graphicx}
\usepackage{bm}
\usepackage{epstopdf}

\newcommand{\eps}{\varepsilon}

\newcommand{\ga}{\gamma}

\newcommand{\om}{\omega}
\newcommand{\prt}{\partial}

\begin{document}

\title{
Dynamics of ring dark solitons in Bose-Einstein condensates and nonlinear optics}

\author{A.M. Kamchatnov}
\ead{kamch@isan.troitsk.ru}
\author{S.V. Korneev}
\ead{svyatoslav.korneev@gmail.com}

\address{
Institute of Spectroscopy, Russian Academy of Sciences, Troitsk,
Moscow Region, 142190, Russia }

\date{\today}

\begin{abstract}
Quasiparticle approach to dynamics of dark solitons is applied to the case of ring solitons.
It is shown that the energy conservation law provides the effective equations of motion of
ring dark solitons for general form of the nonlinear term in the generalized nonlinear
Schr\"odinger or Gross-Pitaevskii equation. Analytical theory is illustrated by examples of
dynamics of ring solitons in light beams propagating through a photorefractive medium and
in non-uniform condensates confined in axially symmetric traps. Analytical results agree
very well with the results of our numerical simulations.
\end{abstract}



\maketitle

\section{Introduction}

As is known, dark solitons are fundamental excitations of nonlinear media with modulationally stable
background (see review articles \cite{kivshar-98,frantz-10} and references therein). Usually their
dynamics is modeled by a generalized nonlinear Schr\"odinger (NLS) or Gross-Pitaevskii (GP) equation
with defocusing (repulsive) nonlinearity
\begin{equation}\label{1-1}
    i\psi_t+\tfrac12\Delta\psi-f(|\psi|^2)\psi=U(\mathbf{r})\psi,
\end{equation}
where $\psi=\psi(\mathbf{r},t)$ is the field variable which interpretation depends of the physical
system under consideration. For example, in case of atomic Bose-Einstein condensate (BEC) it
represents the order parameter (``condensate wave function'') and then $\rho=|\psi|^2$ has a
meaning of density of atoms, $\psi=\sqrt{\rho}\,\exp(i\phi)$, $\mathbf{u}=\nabla\phi$ is the condensate's
flow velocity, and $U(\mathbf{r})$ is the external potential of forces acting on atoms (e.g., the
trap potential). In nonlinear optics, the fields variable $\psi$ has a meaning of the light
field strength, $|\psi|^2$ its intensity, $t$ plays the role of the coordinate along the light
beam, $\mathbf{r}$ is the radius-vector in transverse directions, and $U(\mathbf{r})$ can be
related with non-uniformity of a refractive index.

In the most common case the function
\begin{equation}\label{2-1}
    f(|\psi|^2)=|\psi|^2
\end{equation}
corresponds to a repulsive interatomic interaction in BEC case or a defocusing Kerr nonlinearity
in nonlinear optics case. If $\psi$ depends on time and one space coordinate $x$ only and the medium
is uniform ($U(\mathbf{r})=0$), then Eq.~(\ref{1-1}) with nonlinear term (\ref{2-1}) reduces to
standard 1D NLS equation
\begin{equation}\label{2-2}
    i\psi_t+\tfrac12\psi_{xx}-|\psi|^2\psi=0
\end{equation}
with well-known dark soliton solution
\begin{equation}\label{2-3}
    \psi=\psi_s(x-Vt)=\left\{\sqrt{\rho_0-V^2}\tanh\left[\sqrt{\rho_0-V^2}\,(x-Vt)\right]+iV\right\}
    e^{-i\rho_0t},
\end{equation}
where $\rho_0$ is the background density and $V$ the soliton's velocity. Naturally, the parameters
$\rho_0$ and $V$ are constant here.

In many situations we can approximate locally the soliton solution of the general equation (\ref{1-1})
the soliton solution analogous to formula (4) but now with slow dependence of $\rho_0$ and 
$V$ on space and time coordinates.
For example, if a dark soliton propagates in BEC confined in a cigar-shaped trap, then its dynamics
can be described in certain approximation by  the following generalization of Eq.~(\ref{2-2}):
\begin{equation}\label{2-4}
    i\psi_t+\tfrac12\psi_{xx}-|\psi|^2\psi=U(x)\psi,
\end{equation}
where $U(x)=\omega_0^2x^2/2$ in case of a harmonic axial potential. As was found in \cite{ba2000},
such a soliton oscillates in a harmonic trap with the frequency $\omega=\omega_0/\sqrt{2}$, and
it was shown in \cite{kp04} that this result can be obtained most simply if one considers a
dark soliton as a quasiparticle with the energy
\begin{equation}\label{3-1}
    \eps(V)=\tfrac43(\rho_0-V^2)^{3/2}
\end{equation}
which expression follows easily from the Hamiltonian form of Eq.~(\ref{2-2}). Then conservation of
energy
\begin{equation}\label{3-2}
    \eps(V)=\mathrm{const}
\end{equation}
combined with the Thomas-Fermi (TF) distribution of the background density
\begin{equation}\label{3-3}
    \rho_0(x)=\mu-U(x),\quad \mu=\mathrm{const},
\end{equation}
yields at once the equation for $V=dX/dt$, where $X(t)$ is a coordinate of ``position'' of the soliton:
\begin{equation}\label{3-4}
    \left(\frac{dX}{dt}\right)^2+U(x)=\mathrm{const}.
\end{equation}
This approach was developed in more detail in Ref.~\cite{bkp06} where, in particular, its generality
was illustrated by application to the case of nonlinearity $f(\rho)=\rho^2$. General formulae for the case
of arbitrary function $f(\rho)$ (provided that the condition of modulation stability of the background is
satisfied) were derived later in Ref.~\cite{ks09}. It is clear that this approach is correct as long
as width $\sim1/\sqrt{\rho_0-V^2}$ of the soliton is much less than a characteristic length at which
the background parameters change considerably so that a ``collective'' coordinate $X(t)$ is a
well-defined variable.

This approach admits also the form of Hamilton equations which describe dynamics of the soliton quasiparticle. Indeed, as was shown in \cite{kk-95} (see also \cite{bkp06}), the momentum of the soliton (\ref{2-3})
is equal to
\begin{equation}\label{3a-1}
    p=-2V\sqrt{\rho_0-V^2}+2\rho_0\arctan(\sqrt{\rho_0-V^2}/V)
\end{equation}
and the equation
\begin{equation}\label{3a-2}
    \frac{\prt\eps}{\prt p}=V=\frac{dX}{dt}
\end{equation}
holds which evidently is one of the Hamilton equations for the quasiparticle with the Hamiltonian
(\ref{3-1}). In case of the dark soliton moving through non-uniform BEC confined in a trap created by
the potential $U(x)$ the density $\rho_0$ is a function of $x$ related with $U(x)$ by Eq.~(\ref{3-3}).
Hence the Hamiltonian becomes a function of the position $X$ of the soliton,
$\eps(X,V)\equiv\eps(\rho_0(X),V)$, where $\rho_0(X)$ is related with the potential $U(X)$ by the
equation (\ref{3-3}) with $x$ replaced by the coordinate $X$ of the soliton's position. It is clear
that the equation (\ref{3a-2}) does not change provided the derivative in the left-hand side is
taken for fixed $X$. Then the second Hamilton equation
\begin{equation}\label{3a-3}
    \frac{dp}{dt}=-\left(\frac{\prt\eps(X,V)}{\prt X}\right)_p
\end{equation}
can be readily reduced to
\begin{equation}\label{3a-4}
    2\frac{dV}{dt}=-\frac{\prt U(X)}{\prt X}
\end{equation}
that is to the Newton equation with the effective mass of the dark soliton quasiparticle
equal to the doubled atomic mass as it was first discovered in \cite{ba2000}
(see also \cite{kp04,bkp06,ks09}).

In this Letter we show that the approach of Refs~\cite{kp04,bkp06,ks09} can be easily generalized to
multi-dimensional situations of ring or spherical solitons. For example, if a ring dark soliton
evolves in a nonlinear medium, then
the speed of its radius $R(t)$, $V=dR/dt$, depends not only on time $t$, but also on $R$. If
the width of the soliton is much less than its radius,
then the conservation law of the total energy in the first approximation reads
\begin{equation}\label{3-5}
    2\pi R\eps(\rho_0(R),dR/dt)=\mathrm{const}.
\end{equation}
Thus, we arrive at a differential equation for $R(t)$ which can be easily solved in various situations.
We shall consider here by this method several typical examples. Naturally, in the simplest
case of Kerr nonlinearity and uniform background our approach reproduces well-known results of
Ref.~\cite{ky94}. In actual optics experiments (see, e.g., \cite{exp02}) on ring solitons, the photorefractive
media with $f(\rho)=\rho/(1+\gamma \rho)$ were often used, and in the next section we shall consider evolution
of ring dark solitons in such a medium. Evolution of ring dark solitons in BEC confined in axially
symmetric traps was considered in Refs.~\cite{tfkmk03,tsokf05} in various approximations. We show in
section 3 that our approach gives a very good approximation (different from ones discussed earlier) which
agrees quite well with direct numerical solution of the GP equation up to the moment of decay of ring
solitons to vortices due to their snake instability.

\section{Dynamics of dark ring solitons in a uniform background}

First, we shall consider the case of uniform background which density (intensity)
is constant, $\rho_0=\mathrm{const}$.
Although new results will be obtained for a non-Kerr nonlinearity function $f(\rho)$, we shall
begin with the Kerr-like case $f(\rho)=\rho$ which illustrates very clearly the method.

\subsection{Kerr-like nonlinearity}

In case of Kerr-like nonlinearity (\ref{2-1}), i.e. $f(\rho)=\rho$, the energy of a dark soliton
per unit of its length is given by Eq.~(\ref{3-1}). Hence the total energy of a ring dark soliton
with radius $R$ is equal to $2\pi R\eps$ and its conservation law yields the equation
\begin{equation}\label{4-2}
    R(\rho_0-\dot{R}^2)^{3/2}=R_0(\rho_0-V_0^2)^{3/2},
\end{equation}
where $\dot{R}=dR/dt$ and $R_0$ and $V_0$ are the initial radius and velocity of the dark
soliton, correspondingly.
This equation was derived first in a different form in \cite{ky94} with the use of perturbation
theory. Its differentiation with respect to time $t$ gives the Newton-like equation
\begin{equation}\label{4-3}
    \ddot{R}=\frac1{3R}(\rho_0-\dot{R}^2)=\frac{R_0^{2/3}(\rho_0-V_0^2)}{3R^{5/3}},
\end{equation}
which has been recently obtained in \cite{mss10} as a particular case of equations of contour dynamics
of dark solitons. Variable $R$ can be excluded from (\ref{4-3}) with the help of Eq.~(\ref{4-2})
so that one obtains
\begin{equation}\label{5-1}
    \frac{dV}{dt}=\frac{(\rho_0-V^2)^{5/2}}{3R_0(\rho_0-V_0^2)^{3/2}},
\end{equation}
and elementary integration of this equation yields implicit dependence of $V$ on time
\begin{equation}\label{5-2}
    \frac{V(3\rho_0 -2V^2)}{(\rho_0-V^2)^{3/2}}=\frac{V_0(3\rho_0-2V_0^2)-t/R_0}{(\rho_0-V^2)^{3/2}}.
\end{equation}
This equation can be solved with respect to $V^2$ in a closed form (see \cite{mss10}) and this
analytical solution agrees very well with the results of direct numerical solution of the GP
equation. Thus, interpretation of Eq.~(\ref{4-2}) as the energy conservation law permits us
to reproduce very simply the known results.

\subsection{General form of nonlinearity}

As was shown in \cite{ks09}, in general case the energy of a dark soliton per its unit length can be expressed as
\begin{equation}\label{5-3}
    \eps=\frac12\int_{\rho_m}^{\rho_0}\frac{Q_0(\rho)}{\rho\sqrt{ Q(\rho)}}d\rho,
\end{equation}
where
\begin{equation}\label{5-4}
    Q_0(\rho)=8\rho\int_{\rho}^{\rho_0}[f(\rho_0)-f(\rho')]d\rho',
\end{equation}
\begin{equation}\label{5-5}
    Q(\rho)=Q_0(\rho)-4V^2(\rho_0-\rho)^2,
\end{equation}
and the minimal value $\rho_m$ in the distribution of the density $\rho$ across the dark soliton dip
is related with its velocity $V$ by the equation
\begin{equation}\label{5-6}
    V^2=\frac{Q_0(\rho_m)}{4(\rho_0-\rho_m)^2}.
\end{equation}
Hence, $\eps$ can be considered as a function of $\rho_0$ and $V^2$ and, as above,
conservation of the total energy of the ring soliton reads
\begin{equation}\label{6-1}
    R\eps(\rho_0,\dot{R}^2)=R_0\eps(\rho_0,V_0^2).
\end{equation}
This is a differential equation for the function $R=R(t)$ with given initial conditions
$R=R_0$, $V=V_0$ at $t=0$, and its solution describes the evolution of the ring dark
soliton. To obtain more definite results, one should specify the nonlinearity function $f(\rho)$.

\subsection{Evolution of ring dark solitons in media with photorefractive nonlinearity}

In nonlinear optics experiments \cite{exp02} the nonlinear effects can be modeled by the function
\begin{equation}\label{6-2}
    f(\rho)=\frac{\rho}{1+\gamma\rho}
\end{equation}
which takes into account saturation of the nonlinear response, i.e. $f(\rho)\to1/\gamma=\mathrm{const}$
as $\rho\to\infty$. In this case
\begin{equation}\label{6-3}
    Q_0(\rho)=\frac{8\rho}{\ga^2}\left(\ln\frac{1+\ga\rho_0}{1+\ga\rho}
    +\frac{\ga(\rho_0-\rho)}{1+\ga\rho_0}\right).
\end{equation}
Substitution of this expression into Eqs.~(\ref{5-3}) and (\ref{5-6}) yields the function
$\eps(\rho_0,V^2)$ and the resulting differential equation (\ref{6-1}) can be solved numerically.
Example of such a solution is shown in Fig.~1 for the choice of parameters: $\rho_0=1$, $\ga=1$,
$R_0=30$, $V_0=0.2$. The solid line represents the results of our approximate theory and they
are compared with the exact numerical solution of the generalized NLS equation (\ref{1-1}) with $f$
given by Eq.~(\ref{6-2}) and $U=0$; values of the dark soliton radius for different moments
of time are shown by crosses. As we see, agreement is quite good.
\begin{figure}[bt]
\begin{center}
\includegraphics[width=8cm,height=6cm,clip]{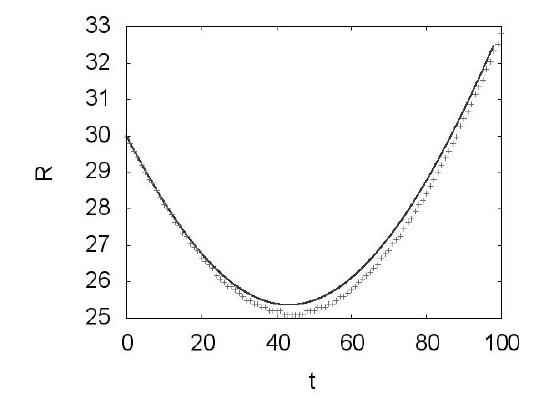}
\caption{Dependence of the radius of the ring dark soliton on time for the case of photorefractive nonlinearity:
solid line corresponds to analytical theory and crosses to numerical solution of the generalized NLS equation.
 }
\end{center}\label{fig1}
\end{figure}

\section{Evolution of ring dark solitons in a non-uniform background}

As a typical example, we shall consider the BEC confined in an axially symmetric trap. As above,
we assume that cylindrical symmetry persists during evolution so that dynamics of a ring dark
soliton reduces to varying of its radius $R$ with time $t$. However, now the background density
$\rho_0$ is not uniform and depends on the radial coordinate $r$. In the Thomas-Fermi (TF) approximation
and for quadratic nonlinearity $f(\rho)=\rho$ this dependence is given by
\begin{equation}\label{7-1}
    \rho_0(r)=\mu-U(r)
\end{equation}
and in the case of a harmonic trap $U(r)=\omega_0^2r^2/2$ this dependence can be written as
\begin{equation}\label{7-2}
    \rho_0(r)=\tfrac12\omega_0^2(l^2-r^2),
\end{equation}
where $l$ is the radius of the condensate's distribution related with the chemical potential $\mu$
and the number of particles $N$ (per unit axial length) by the formulae
\begin{equation}\label{7-3}
    \mu=\tfrac12\omega_0^2l^2,\qquad N=\frac{\pi}4\omega_0^2l^4.
\end{equation}
Therefore the energy conservation law now reads
\begin{equation}\label{7-4}
    R\eps(\rho_0(R),\dot{R}^2)=R_0\eps(\rho_0(R_0),V_0^2).
\end{equation}
We shall study it in some detail for a particular case of a harmonic trap when $\rho_0$ is given by
Eq.~(\ref{7-2}) and of two-particle repulsive interatomic interaction when $f(\rho)=\rho$ and
$\eps(V)$ is given by Eq.~(\ref{3-1}). Then Eq.~(\ref{7-4}) takes the form
\begin{equation}\label{7-5}
    R\left[\frac{\omega_0}2(l^2-R^2)-\left(\frac{dR}{dt}\right)^2\right]^{3/2}=
    R_0\left(\rho_0-V_0^2\right)^{3/2},
\end{equation}
where $\rho_0=\omega_0^2(l^2-R_0^2)/2$ is the density at the initial radius $R_0$ of the soliton.
This equation can be rewritten as
\begin{equation}\label{8-1}
    \left(\frac{dR}{dt}\right)^2=\tfrac12\omega_0^2(l^2-R^2)-\frac{R_0^{2/3}(\rho_0-V_0^2)}{R^{2/3}}.
\end{equation}
Its differentiation with respect to time $t$ gives the Newton-like equation
\begin{equation}\label{8-2}
    \frac{d^2R}{dt^2}=-\frac{\omega_0^2}2R+\frac{R_0^{2/3}(\rho_0-V_0^2)}{3R^{5/3}}.
\end{equation}
The first term in the right-hand side corresponds to oscillations of the soliton radius with the frequency
$\omega_0/\sqrt{2}$, i.e. it describes the effects of non-uniformity of the condensate discovered
in \cite{ba2000}. The second term in the right-hand side of (\ref{8-2}) describes the effects of the curvature
of the soliton and coincides with Eq.~(\ref{4-3}). In our adiabatic approximation both effects
are weak and therefore they are added in the combined equation (\ref{8-2}). Integration of this
equation reduces to a quadrature
\begin{equation}\label{8-3}
    \frac{\omega_0t}{\sqrt{2}}=\pm\int_{R_0/l}^{R/l}\frac{dw}{\sqrt{1-w^2-aw^{-2/3}}},
    \quad a=\frac{2R_0^{2/3}(\rho_0-V_0^2)}{\omega_0^2l^{8/3}},
\end{equation}
which actually solves the problem.
\begin{figure}[bt]
\begin{center}
\includegraphics[width=8cm,height=6cm,clip]{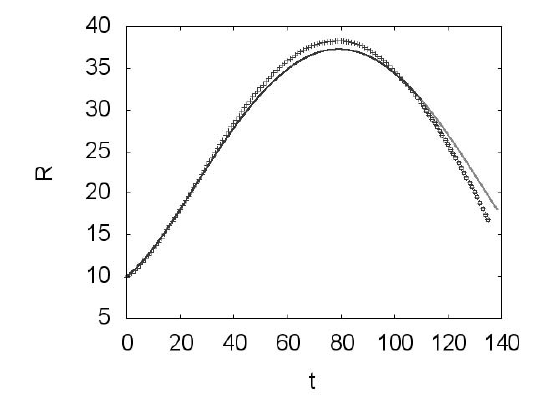}
\caption{Dependence of the radius of the ring dark soliton on time in case of its propagation along
TF distribution of density BEC confined in a harmonic trap:
solid line corresponds to analytical theory and crosses to numerical solution of the GP equation.
 }
\end{center}\label{fig2}
\end{figure}
In Fig.~2 we have compared this analytical solution with the exact numerical solution of Eq.~(\ref{1-1})
for the choice of the parameters $N=3486.5$, $\mu=1$, $l=50$, $\om_0=0.03$, $R_0=10.0$, $V=0.229$.
At the moment of time $t_c=79.3$ the ring soliton reaches its turning point with the maximal value of
the radius $R_{max}=37.25$; after this moment its radius decreases and, as it was already indicated in
Ref.~\cite{tfkmk03}, the snake instability starts to develop, so that the dark soliton evolves into
necklace of vortices. This process is illustrated in Fig.~3.
\begin{figure}[bt]
\begin{center}
\includegraphics[width=6cm,height=6cm,clip]{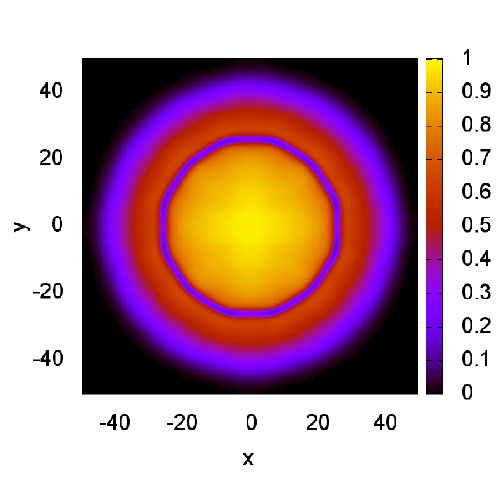}
\includegraphics[width=6cm,height=6cm,clip]{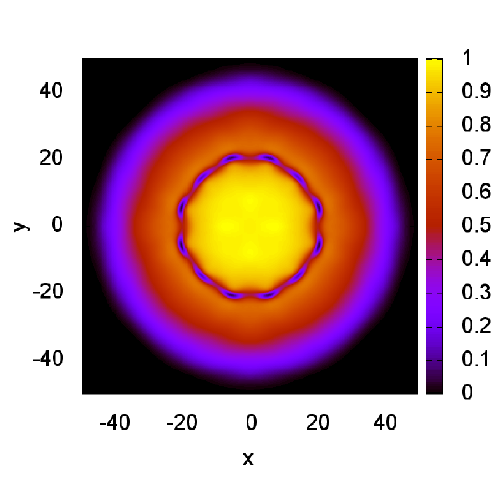}
\caption{Transformation of a ring dark soliton into necklace of vortices: density plots at the moments
of time (a) $t=120 $ with a slightly disturbed dark soliton and (b) $t=130 $ with well developed necklace
of vortices.
 }
\end{center}\label{fig3}
\end{figure}
In Fig.~2 the radius of the necklace is shown by circles and formal analytical solution corresponds here
to a dashed line. As a measure of deviation of the structure from the ring soliton we have introduced
the quantity
\begin{equation}\label{9-1}
    \delta\rho=\sqrt{\overline{\rho^2}-(\overline{\rho})^2},
\end{equation}
where the averages $\overline{\rho}$ and $\overline{\rho^2}$ are taken along the circle of the minimal
values of the density in the structure. Its dependence on time is shown in Fig.~4.
\begin{figure}[bt]
\begin{center}
\includegraphics[width=8cm,height=6cm,clip]{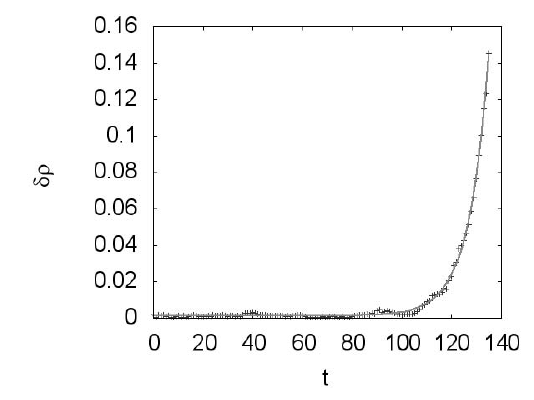}
\caption{Dependence of the variable $\delta\rho$ on time: values from numerical solution of the GP
equation are shown by crosses; a dashed line is a fitting curve (\ref{9-2}).
 }
\end{center}\label{fig4}
\end{figure}
It can be fitted by the exponential growth of $\delta\rho$:
\begin{equation}\label{9-2}
    \delta\rho\approx 8.198\cdot 10^{-9}\exp(0.123 t)
\end{equation}
The coefficient can be represented as $8.198\cdot 10^{-9}=0.0065\cdot\exp(-0.123\cdot 110)$
which corresponds to the numerically introduced disturbance about $\sim 0.01$ which starts to
grow up after the moment of time about $t\approx110$, that is after the turning point
what can be explained as a result of very slow change of parameters of the soliton in
vicinity of the turning point so that the most unstable mode has enough time for its
evolution without strong competition with other modes.
The growth rate constant $\Gamma=0.123$ can be compared with that calculated according to the theory
\cite{kt-1988} of snake instability of plane dark solitons for given depth and background density.
If we take these parameters for the moment $t=110$, when the soliton radius equals to $R=30.9$, then
we get the maximal growth rate $\Gamma=0.09$ for the unstable mode with the wave number $k=0.39$
in qualitative agreement with the numerically found value.
It is worth noting that we have here the number of wavelengths along the ring of the soliton
equal to $Rk\cong 12$ in agreement with the number vortices formed at the nonlinear stage of
development of the instability. Thus, the snake instability theory agrees at least qualitatively
with the process observed in numerical simulations.

\section{Conclusion}

The idea that bright solitons behave as ``particles'' goes back to the beginning of modern nonlinear
physics. However, it was not so obvious that it could be applied to dark solitons since their
behavior depends also on the background evolution. Konotop and Pitaevskii noticed \cite{kp04} that if
background is a stationary solution of the GP equation in the TF approximation, then dynamics of
dark solitons is determined by conservation of their energy.
We have shown in this Letter that this idea can be extended to dynamics of ring solitons.
This simple approach can be easily generalized to arbitrary form of the nonlinear repulsive
self-action what is important for applications to nonlinear optics. Examples discussed in
this Letter show that this approach gives accurate enough description of motion of ring dark solitons
up to the moment of their decay to vortices due to snake instability.

\subsection*{Acknowledgments}

We are grateful to Yu.G.~Gladush and L.A.~Smirnov for useful discussions.
This work was supported by RFBR (grant 09-02-00499).
We thank also Joint Supercomputer Center of RAS for the provision of computing facilities.

\end{document}